\documentclass{svjour2}                    % onecolumn
\smartqed  % flush right qed marks, e.g. at end of proof
\usepackage{graphicx}
\begin{document}

\title{An investigation of chaotic diffusion in a family of Hamiltonian 
mappings whose angles diverge in the limit of vanishingly action}

\titlerunning{Chaotic diffusion in a family of maps $\ldots$}

\author{$^{1,2}$Edson D.\ Leonel \and
        $^{1}$C\'elia M.\ Kuwana
}

%\authorrunning{Short form of author list} % if too long for running head

\institute{$^1$UNESP - Univ Estadual Paulista, Departamento de F\'isica \at
              Av. 24A 1515 Rio Claro - SP - Brazil\\
              Tel.: +55-19-3526 9174\\
              Fax: +55-19 3526 9181\\
              \email{edleonel@rc.unesp.br}           %  \\
%             \emph{Present address:} of F. Author  %  if needed
           \and
           $^2$Abdus Salam International Center for Theoretical Physics, 
Strada Costiera 11, 34151 Trieste, Italy\\
}

\date{Received: date / Accepted: date}
% The correct dates will be entered by the editor

\maketitle

\begin{abstract}
The chaotic diffusion for a family of Hamiltonian mappings whose 
angles diverge in the limit of vanishingly action is investigated by 
using the solution of the diffusion equation. The system is described 
by a two-dimensional mapping for the variables action, $I$, and angle, 
$\theta$ and controlled by two control parameters: (i) $\epsilon$, 
controlling the nonlinearity of the system, particularly a transition 
from integrable for $\epsilon=0$ to non-integrable for $\epsilon\ne0$ 
and; (ii) $\gamma$ denoting the power of the action in the equation 
defining the angle. For $\epsilon\ne0$ the phase space is mixed and 
chaos is present in the system leading to a finite diffusion in the 
action characterized by the solution of the diffusion equation. The 
analytical solution is then compared to the numerical simulations 
showing a remarkable agreement between the two procedures.

% \subclass{MSC code1 \and MSC code2 \and more}
\end{abstract}

\section{Introduction}
\label{sec2}

The understanding of diffusive process has intrigued the humanity for 
long while. The reasons are vast and include since a very simple drop 
of colored ink in a jar with a fluid \cite{p1}, passing from other 
fields including medicine \cite{p2} such as how a given drug diffusing 
in the blood reaches a certain vital organ, biology/ecology 
\cite{p3} where pollen from a plant diffuses to reach another plant, 
environmental sciences with diffusion of pollution, no matter 
if gas (air) \cite{p4} with large impact in the earth, fluctuating 
solids in oceans traveling over the continents \cite{b1}, water 
percolation \cite{p5} in the ground transporting chemical reagents 
from pesticides to the water table and many other areas. In physics and 
considering its importance to the area, the subject is a so standard 
topic \cite{b2,b3,b4} which is delivered in undergrad courses around 
the world where basic properties of random walk are introduced together 
with other approaches.

In this paper we investigate the chaotic diffusion in a family of 2-D 
area preserving mappings described in the variables angle and action. 
The action is controlled by a parameter $\epsilon$ controlling also the 
nonlinearity of the system and particularly a transition from 
integrability to non integrability \cite{b5}. If $\epsilon=0$ the 
system is integrable and the phase space is regular. For $\epsilon\ne 
0$, a mixed phase space is produced with the coexistence of regions with 
regularity marked by periodic islands and invariant spanning curves 
limiting the size of a chaotic sea, hence limiting the diffusion of 
chaotic particles. The variable angle is defined in such a way it 
diverges in the limit of vanishingly action \cite{p6}. The speed of the 
divergence is controlled by a parameter $\gamma>0$. Statistical 
properties of the chaotic sea have already been discussed under 
different approaches \cite{p7,p8,p9,p10}, mainly considering 
phenomenological techniques and numerical simulations. Scaling laws 
\cite{p11} leading to scaling invariance of the chaotic 
sea furnish an overall characterization of the universality of the 
problem. Our main goal in this paper is to describe the behavior of the 
statistical properties of the chaotic sea by using so far a solution of 
the diffusion equation \cite{b4} under specific boundary conditions. 
Our results are then compared to those already known in the literature 
\cite{p11} showing a remarkable agreement of the procedure.

This paper is organized as follows. In Sec. \ref{sec3} we describe the
Hamiltonian and the family of mappings showing applications for different
systems. The scaling properties present in the system are 
highlighted also. Section \ref{sec4} is devoted to discuss the solution 
of the diffusion equation and its implications to the dynamics for 
different time scalings as well as initial conditions. Our discussions 
and conclusions are presented in Sec. \ref{sec5}.

\section{The mapping and its properties}
\label{sec3}

The dynamics of an autonomous two degrees of freedom system can be 
described by a generic Hamiltonian \cite{b5} as 
$H(I_1,I_2,\theta_1,\theta_2)=H_0(I_1,I_2)+\epsilon 
H_1(I_1,I_2,\theta_1,\theta_2)$ where the term $H_0(I_1,I_2)$ gives the 
integrable part while $H_1(I_1,I_2,\theta_1,\theta_2)$ contributes to 
the non integrable part. The parameter $\epsilon$ controls a transition 
from integrability with $\epsilon=0$ to non-integrability for 
$\epsilon\ne0$. Given the energy of the system is constant, the variable 
$I_2$ can be eliminated lasting three relevant variables. Using a 
Poincar\'e section in the plane $I_1\times\theta_1$ with $\theta_2$ 
constant the 3-D flow is reduced to an application in a 2-D mapping in 
the plane. A most generic mapping \cite{b5} that describes the dynamics 
is given by
\begin{equation}
\left\{\begin{array}{ll}
I_{n+1}=I_n+\epsilon h(\theta_n,I_{n+1})\\
\theta_{n+1}=[\theta_n+K(I_{n+1})+\epsilon p(\theta_n,I_{n+1})]~~{\rm mod
(2\pi)}\\
\end{array}
\right.,
\label{eq2}
\end{equation}
where $h(\theta_n,I_{n+1})$, $K(I_{n+1})$ and $p(\theta_n,I_{n+1})$ are
nonlinear functions of their variables. The integer $n$ denotes the 
iterated of the mapping which preserves the area only if
${{\partial p(\theta_n,I_{n+1})}\over{\partial \theta_n}}+{{\partial
h(\theta_n,I_{n+1})}\over{\partial I_{n+1}}}=0$.

Using $p(\theta_n,I_{n+1})=0$ and $h(\theta_n)=\sin(\theta_n)$, 
different applications were already considered in the literature and to 
mention few of them
\begin{itemize}
\item{$K(I_{n+1})=I_{n+1}$, describing the Chirikov-Taylor standard 
mapping \cite{JP31};
}
\item{$K(I_{n+1})=2/I_{n+1}$, gives the dynamics of a Fermi-Ulam model
\cite{JP32,JP33};
}
\item{$K(I_{n+1})=\zeta I_{n+1}$, where $\zeta$ is a constant describes 
a bouncer model \cite{JP34};
}
\item{While for the case
$$
K(I_{n+1})=\left\{\begin{array}{ll}
4\zeta^2(I_{n+1}-\sqrt{I_{n+1}^2-{{1}\over{\zeta^2}}})~~{\rm
if}~~I_{n+1}>{{1}\over{\zeta}},\nonumber\\
4\zeta^2 I_{n+1}~~{\rm if}~~I_{n+1}\le{{1}\over{\zeta}}.
\nonumber
\end{array}
\right.
$$
with $\zeta$ constant recovers the hybrid Fermi-Ulam bouncer model
\cite{JP35,EDL1,EDL2};
}
\item{$K(I_{n+1})=I_{n+1}+\zeta I_{n+1}^2$ recovers the logistic twist 
mapping \cite{JP30}.
}
\end{itemize}

\begin{figure}[t]
%\vspace*{-0.8cm}
\centerline{\includegraphics[width=1.0\linewidth]{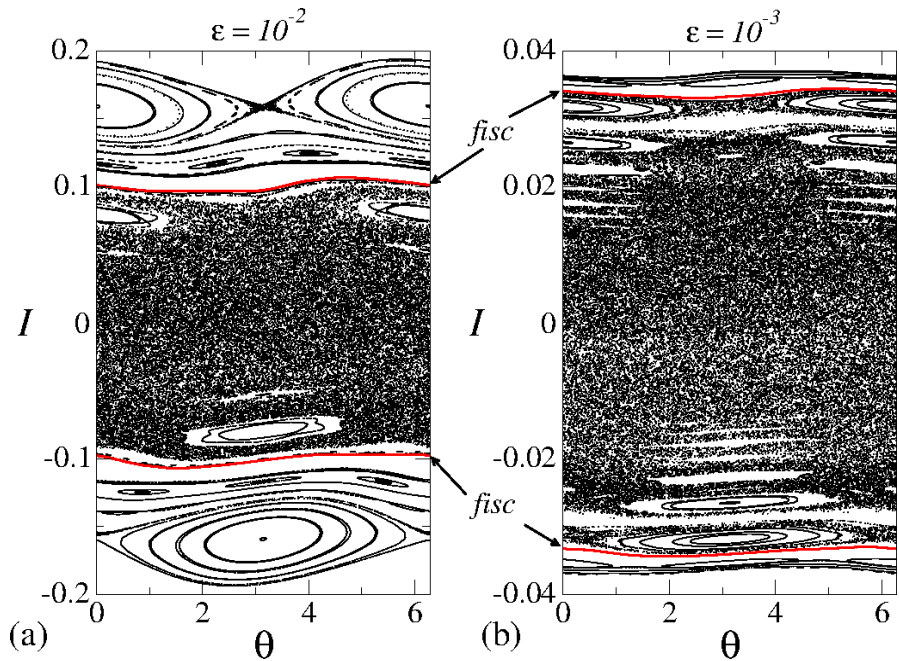}}
\caption{{Plot of the phase space for mapping (\ref{eq4}) considering 
$\gamma=1$ and: (a) $\epsilon=10^{-2}$; (b) $\epsilon=10^{-3}$. The 
invariant spanning curves, represented by {\it fisc} are shown in red 
(gray) color.}}
\label{Fig1}
\end{figure}

Our goal in this paper is to investigate the dynamical properties 
for chaotic orbits considering a family of mappings described by
$h(\theta_n,I_{n+1})=\sin(\theta_n)$ and $K=1/|I_{n+1}|^{\gamma}$ with 
$\gamma>0$ and $p(\theta_n,I_{n+1})=0$, leading to
\begin{equation}
\left\{\begin{array}{ll}
I_{n+1}=I_n+\epsilon \sin(\theta_n)\\
\theta_{n+1}=[\theta_n+{{1}\over{|I_{n+1}|^{\gamma}}}]~~{\rm mod 
(2\pi)}\\
\end{array}
\right..
\label{eq4}
\end{equation}
We see that $\epsilon=0$ leads the phase space to be regular with 
$I$ constant while for 
$\epsilon\ne0$ the regularity is partially destroyed being replaced by 
a mixed phase space including coexistence of chaos and regularity marked 
by periodic islands and invariant spanning curves limiting the size of
the chaotic sea. Figure \ref{Fig1} shows a plot of the phase space 
of mapping (\ref{eq4}) for the parameters $\gamma=1$ and: (a) 
$\epsilon=10^{-2}$ and (b) $\epsilon=10^{-3}$. The invariant 
spanning curves are shown in red (gray) color. Other values for the 
control parameters produce similar plots\footnote{Similar in the sense 
of mixed with coexistence of chaos and regularity including periodic 
islands and invariant spanning curves.}.

The chaotic sea is produced by absence of correlations between 
$\theta_{n+1}$ and $\theta_n$ in the limit of $I$ sufficiently small. 
For large values of the action correlations appear in $\theta$ 
producing regularity to the phase space consequently creating the 
invariant spanning curves, which play a major role in the dynamics 
\cite{p7}. The lowest one, denominated here as $I_{fisc}$, in either 
positive and negative sides work as a boundary prohibiting transport of 
particles through them. The index $fisc$ is a denotation to first 
invariant spanning curve. It is known \cite{b5,p8} that the position of 
the first invariant spanning curves from the mapping (\ref{eq4}) can be 
described using a result imported from the Chirikov-Taylor standard map. 
Indeed in a transition from local to globally chaotic dynamics 
\cite{p11}, the position of the first invariant spanning curves are 
given by 
$I_{fisc}=\left[{{\gamma}\over{K_{ef}}}\right]^{{1}\over{\gamma+1}}
\epsilon^{{1}\over{\gamma+1}}$, with a second order correction given by
$\bigcirc^2\left({{\Delta 
I}\over{\tilde{I}}}\right)=-{{1}\over{2}}\left[{{K_{ef}}
\over{\gamma}}\right]^{{1}\over{\gamma+1}}\epsilon^{{\gamma}\over{
\gamma+1}}$. Here $K_{ef}$ denotes an effective parameter which 
describes locally the transition from local to globally chaotic 
dynamics. Starting the dynamics with very low action, a particle 
or an ensemble of non interacting particles, diffuses along the chaotic 
sea. The observable of interest is defined as 
$I_{rms}(n)=\sqrt{\overline{I^2}(n)}$, where 
\begin{equation}
\overline{I^2}={{1}\over{M}}\sum_{i=1}^M\left[{{1}\over{n}}\sum_{j=1}
^nI^2_{i,j}\right],
\label{eq5}
\end{equation}
with $M$ corresponding to an average over an ensemble of 
$\theta\in[0,2\pi]$ and $n$ is the number of iterations of the mapping. 
The summation in $j$ is taken over the orbit while the summation in $i$ 
is realized over the ensemble of initial conditions. Indeed $I_{rms}$ 
behaves as follows. For short $n$ and $I_0\cong0$, $I_{rms}$ starts to 
growth \cite{p11} 
with $n$ as $I_{rms}\propto (n\epsilon^2)^{\beta}$ where $\beta=1/2$. 
This exponent is a confirmation that a chaotic particle diffuses as a 
random walk particle. For large enough $n$, $I_{rms}$ saturates due to 
the limits of the allowed regions to visit in the phase space. At such 
domain, $I_{rms}\propto \epsilon^{\alpha}$ where $\alpha$ is a critical 
exponent \cite{p7}. In fairness, 
$I_{sat}=\lim_{n\rightarrow\infty}I_{rms}$ must scale with the size of 
the chaotic sea. The limitations are $\pm I_{fisc}$, then 
$\alpha={{1}\over{1+\gamma}}$. The changeover from 
growth to the saturation is marked by $n_x\propto\epsilon^z$. The 
relevant scaling law already known \cite{p11} is 
$z={{\alpha}\over{\beta}}-2$, leading to 
$z=-{{2\gamma}\over{\gamma+1}}$.

It the initial action is no longer small enough but is still smaller 
than $I_{sat}$, the ensemble of particles diffuses as follows. Part of 
the ensemble increases action with probability $p$ and part of it 
decreases with probability $q=1-p$. Since there is no bias in the 
system, then $p=q$ and half of the ensemble increases/decreases. 
Eventually this symmetry is broken \cite{p9} and an additional scaling 
is observed. $I_{rms}$ stays in a plateau until a crossover time 
$n_n^{\prime}$ that scales with $n_x^{\prime}\propto 
{{I_0^2}\over{\epsilon^2}}$. If an ensemble of initial conditions is 
given above of the saturation and below of the first invariant spanning 
curve the scenario can be very complicated due to the existence of 
stickiness \cite{p12}. The relevant scaling transformations 
\cite{p8,p11} that lead all the curves of $I_{rms}$ obtained from 
different $\epsilon$ and $I_0$ to overlap each other are: (i) 
$I_{rms}\rightarrow I_{rms}/\epsilon^{\alpha}$; (ii) $n\rightarrow 
n/\epsilon^z$ and; (iii) 
$I_0^{\prime}=I_0\left({{\epsilon^{\prime}}\over{\epsilon}}\right)^{{1}
\over{1+\gamma}}$.

\section{Solution for the diffusion equation}
\label{sec4}

The characteristics observed in the phase space, particularly in the 
chaotic dynamics, allow us to make a connection with diffusive 
problems. Firstly when an initial condition is given in the chaotic sea 
at the regime of low action, typically $I_0\sim 10^{-3}\epsilon$, the 
dynamics diffuses chaotically along the phase space passing near the 
islands and visiting regions close to the invariant spanning curves. 
Because the mapping is area preserving the transport of particles 
through such curves is not allowed. An immediate conclusion is that, 
once in the chaos, always in the chaos.

Whenever making a connection with statistical mechanics, a chaotic 
particle has probability $p>0$ of moving one side, say increasing the 
action, and $q=1-p>0$ of moving other side, decreasing the action,  
therefore resembling a motion of a random walk 
process. Due to the property of area preservation, a particle can not 
traverse the invariant spanning curves. They work as reflecting 
barriers\footnote{We assume them as reflecting boundaries, however a 
short discussion must be made here. When a particle passes near 
enough of a regular region it might suffers a dynamical trapping 
called stickiness. The particle stays confined in such a region for a 
while, that may be eventually very long, until escape such region and 
visit other regions of the phase space. During a temporary trapping 
the diffusion is no longer normal but rather anomalous.}. A diffusion 
equation \cite{b4} written for the variable action $I$ and the number 
of iterations $n$, denoting the time, is given by
\begin{equation}
{{\partial P(I,n)}\over{\partial 
n}}=D{{\partial^2P(I,n)}\over{\partial I^2}},
\label{eq_diff}
\end{equation}
where $D$ is the diffusion coefficient obtained from the relation 
$D={{(\overline{\Delta 
I})^2}\over{2}}=\bar{I^2}_{n+1}-\bar{I^2}_n={{\epsilon^2}\over{4}}$ and 
$P(I,n)$ gives the probability of observe a given action 
$I\in[-I_{fisc},+I_{fisc}]$ at a given time $n$. The boundary 
conditions for this problem are ${{\partial P}\over{\partial 
I}}\big\vert_{\pm I_{fisc}}=0$ implying no flux of particles through the 
invariant spanning curves.

There are many different ways of solving the equation (\ref{eq_diff}). 
In our case we used a technique of separation of variables therefore 
writing $P(I,n)=X(I)N(n)$ where, as usual, $X(I)$ is a function that 
depends only on $I$ and $N(n)$ is another function that depends only 
on $n$. This technique transforms the original partial differential 
equation, first order in $n$ and second order in $I$ into two ordinary 
differential equations that must be solved separately. Such separation 
is allowed under the assumption that the variables $I$ and $n$ are 
independent of each other. We considered also that at a time $n=0$, all 
the initial particles were localized at $P(I,0)=\delta(I-I_0)$ with 
$I_0\in[-I_{fisc},+I_{fisc}]$ and along the chaotic sea. $I_0\cong 0$ 
gives the maximal diffusion possible observed for $I_{rms}$. When 
$I_0\ne 0$ an additional scaling is observed in the dynamics 
\cite{p8,p11}.

Since the procedure is standard in many textbooks, see for example Ref. 
\cite{b4}, we report only the final solution
\begin{equation}
P[I(n)]={{1}\over{2I_{fisc}}}+{{1}\over{I_{fisc}}}\sum_{k=1}^{\infty}
\cos\left[{{k\pi(I-I_0)}\over{I_{fisc}}}\right]e^{-{{k^2\pi^2Dn}\over
{ I^2_{fisc}}}},
\label{solution}
\end{equation}
where $I_0$ defines the initial action along the chaotic sea, $n$ 
corresponds to the number of iterations of the mapping and $k$ comes 
from the boundary conditions and must be in the interval 
$k\in[1,\infty)$. In practical, the summation does not need to run to 
infinity. Because the exponentials decay with the square of $k$, few 
terms on $k$ are enough. We have used for safety in our numerical 
results $k=100$ but one can not see great differences of $k=100$ with 
the first order approximation $k=1$, which is indeed the leading term of 
the summation\footnote{We have also compared the dynamics with 
$k=10$, $k=10^3$ and $k=10^4$ and no difference was noticed.}.

The knowledge of the probability function $P[I(n)]$ allows us to obtain 
different observables in the phase space. Since the phase space is 
symmetrical with respect to the action $I$, particularly the symmetry 
for $I$ inside of the range $I\in[-I_{fisc},+I_{fisc}]$, the most 
interesting observable to be studied is $\overline{I^2}(n)$ instead of 
$\overline{I}(n)$. It is obtained from direct integration of 
$\overline{I^2}(n)=\int_{-I_{fisc}}^{I_{fisc}}I^2P[I(n)]dI$, which 
leads to the following expression
\begin{equation}
\overline{I^2}(n)=I^2_{fisc}\left[{{1}\over{3}}+{{4}\over{\pi^2}}\sum_{k
=1}^{\infty}{{(-1)^k}\over{k^2}}\cos\left({{k\pi 
I_0}\over{I_{fisc}}}\right)e^{-{{k^2\pi^2Dn}\over{I^2_{fisc}}}}
\right].
\label{i2}
\end{equation}
To have a clear comparison of the results produced by Equation 
(\ref{i2}) we have to obtain the observable 
$I_{rms}(n)=\sqrt{\overline{I^2}(n)}$.

\begin{figure}[t]
%\vspace*{-0.8cm}
\centerline{\includegraphics[width=0.8\linewidth]{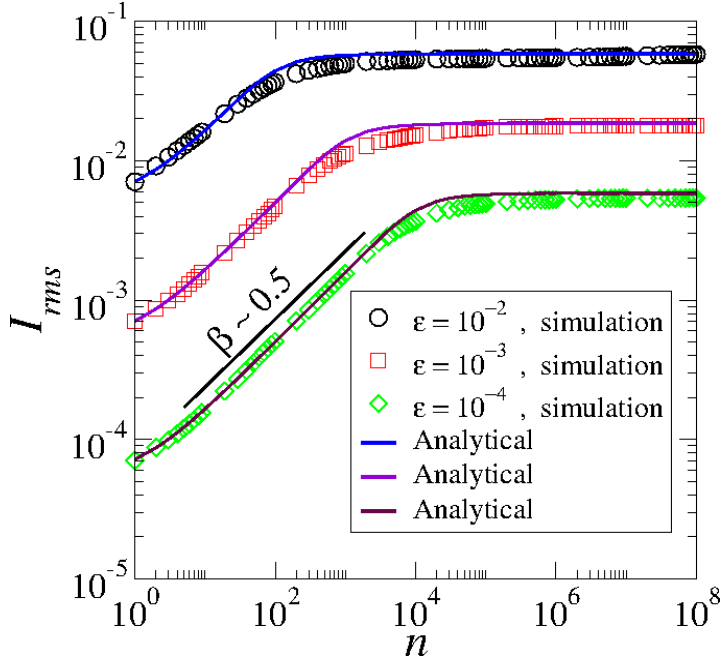}}
\caption{{Plot of $I_{rms}~vs.~n$ for different control parameters as 
labeled in the figure. Symbols are used for numerical simulation while 
continuous line represent the analytical solution given by equation 
(\ref{irms}).}}
\label{Fig2}
\end{figure}

Since Equation (\ref{eq_diff}) was solved to give the probability of 
finding an action $I$ already averaged over an ensemble at an instant 
$n$, then we have to take an average on $n$ in the Equation (\ref{i2}) 
to make comparisons feasible. We notice the sum in $n$ for Equation 
(\ref{i2}) affects only the exponential term. The summation over the 
exponential terms gives a perfect geometrical series which converges 
well. Averaging then Equation (\ref{i2}) over $n$ we end up with the 
following expression
\begin{eqnarray}
&I_{rms}&(n)=I_{fisc}\label{irms}\\
&\times&\sqrt{{{1}\over{3}}+{{4}\over{\pi^2}}
\sum_ { k=1 } ^ { \infty}{{(-1)^k}\over{k^2}}\cos\left({{k\pi 
I_0}\over{I_{fisc}}}\right)e^{-{{k^2\pi^2D}\over{I^2_{fisc}}}}\left[{{1}
\over{n}}\left({{(1-e^{-{{k^2\pi^2Dn}\over{I^2_{fisc}}}})}\over{(1-e^{-{
{k^2\pi^2D}\over{I^2_{fisc}}}} ) } } \right) \right]
},\nonumber
\end{eqnarray}
when $n$ must run from $n=1,2\ldots \infty$. Figure \ref{Fig2}
shows a plot of $I_{rms}(n)~vs.~n$ for $\gamma=1$ and three control 
parameters namely: 
$\epsilon=10^{-2}$, $\epsilon=10^{-3}$ and $\epsilon=10^{-4}$. We used 
$I_0=10^{-3}\epsilon$ as initial action leading to a situation of 
maximizing the diffusion for $I_{rms}$. Symbols represent the numerical 
simulations while continuous line are the theoretical result. The error 
bar produced by the simulations for an ensemble of $M=5\times 10^3$ 
different initial conditions is smaller then the symbols size. We see 
that the curve starts to 
growth with an exponent $\beta$ as a function of $n$ and suddenly it 
bends towards a regime of saturation for large enough $n$. The 
changeover from growth to the saturation is defined by a crossover 
$n_x$. The slope of growth is marked by an exponent $\beta\cong 1/2$ as 
theoretically foreseen in Ref. \cite{p11}. For large enough $n$, the 
saturation of the curves represents the influence of the invariant 
spanning curves in the phase space therefore limiting the diffusion in 
the action. For the regime of $n\rightarrow\infty$, 
$\lim_{n\rightarrow\infty}I_{rms}={{I_{fisc}}\over{\sqrt{3}}}\cong
{{I_{fisc}}\over{1.732\ldots}}$. A naive estimation of $I_{rms}$ in 
Ref. \cite{p11} led to $I_{sat}\sim I_{fisc}/1.8$, therefore furnishing 
a good agreement between the theoretical and numerical results.

Let us now discuss the leading term of equation (\ref{irms}). Whenever 
considering a first order approximation we chose $k=1$ and two 
different cases: (i) $I_0\cong 0$ and (ii) $0<I_0<I_{fisc}/\sqrt(3)$. 
We start first with case (i). The nonlinear function used in equation 
(\ref{irms}) can be expanded in Taylor series until first order as
\begin{eqnarray}
\cos\left({{\pi 
I_0}\over{I_{fisc}}}\right)&\cong&\left[1-{{1}\over{2}}\left({{\pi I_0}
\over{I_{fisc}}}\right)^2\right],
\label{coseno}\\
e^{-{{\pi^2D}\over{I^2_{fisc}}}}&\cong&1-{{\pi^2D}\over{I^2_{fisc}}},\\
1-e^{-{{\pi^2D}\over{I^2_{fisc}}}}&\cong&{{\pi^2D}\over{I^2_{fisc}}}.
\end{eqnarray}
Because of the presence of the factor ${{1}\over{n}}$ in equation 
(\ref{irms}), the expression for the exponential depending on $n$ must 
be expanded to the second order, then
\begin{equation}
1-e^{-{{\pi^2Dn}\over{I^2_{fisc}}}}\cong{{\pi^2D}\over{I^2_{fisc}}}-{{
\pi^4D^2n^2}\over{2I^4_{fisc}}}.
\end{equation}
Substituting these terms in equation (\ref{irms}), grouping them 
properly and after consider that $D={{\epsilon^2}\over{4}}$ we obtain
\begin{equation}
I_{rms}(n)\cong 
\sqrt{\left({{1}\over{3}}-{{4}\over{\pi^2}}\right)\epsilon^{{1}\over{
1+\gamma}}+{{\epsilon^2n}\over{2}}},
\end{equation}
and that the variation in $n$ leads to 
$I_{rms}\propto\sqrt{{\epsilon^2n}\over{2}}$. This result confirms the 
exponent $\beta=1/2$ as well as the transformation ad-hoc $n\rightarrow 
n\epsilon^2$ used in Ref. \cite{p11}. When such a regime of growth 
intersects $I_{sat}$, the crossover emerges, hence 
$\sqrt{{\epsilon^2n_x}\over{2}}={{\epsilon^{{1}\over{1+\gamma}}}\over{
\sqrt{3}}}$, leading to
\begin{equation}
n_x\propto {{2}\over{3}}\epsilon^{-{{2\gamma}\over{1+\gamma}}},
\end{equation}
therefore $z=-{{2\gamma}\over{1+\gamma}}$.

\begin{figure}[t]
%\vspace*{-0.8cm}
\centerline{\includegraphics[width=0.5\linewidth]{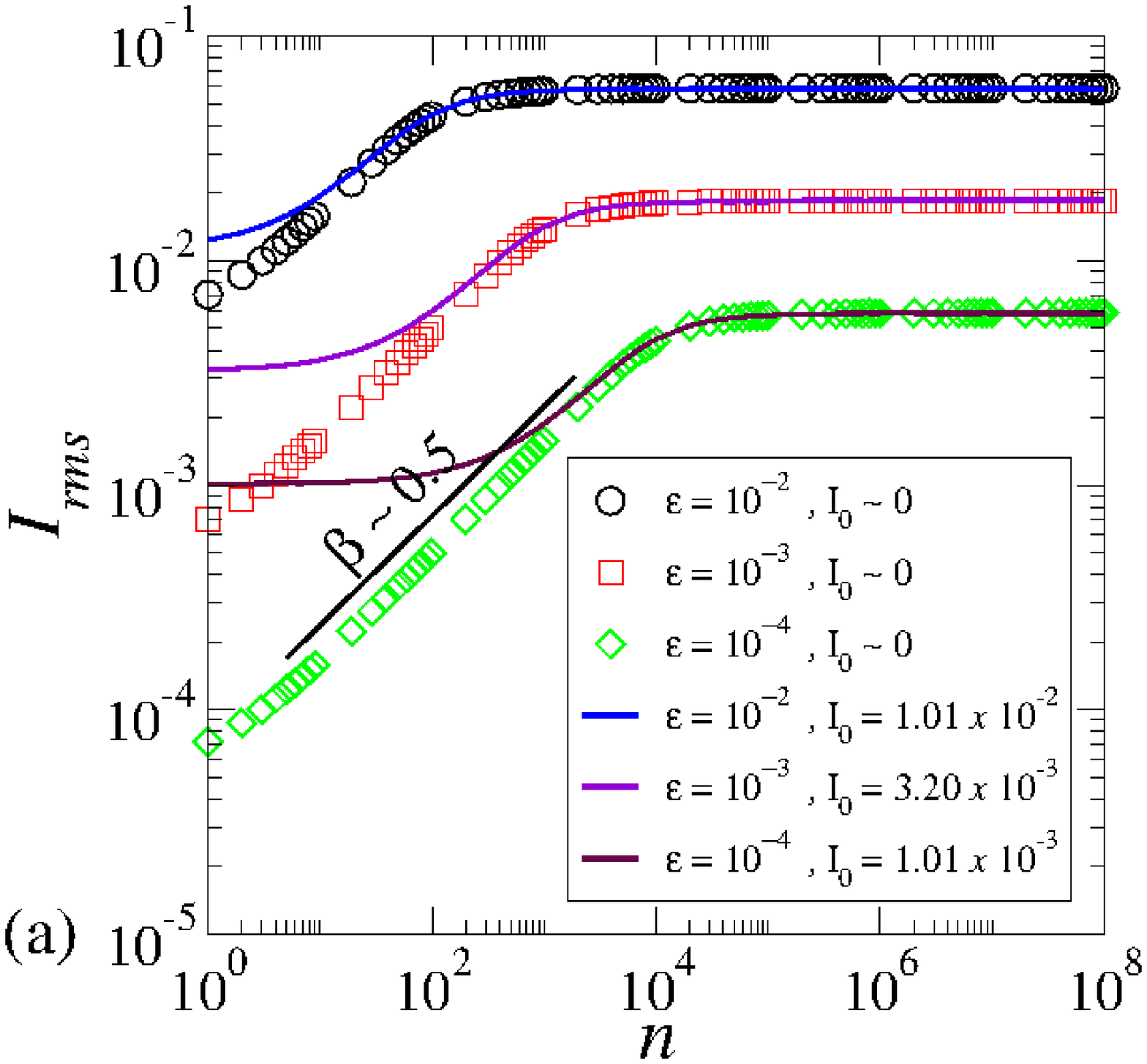}
\includegraphics[width=0.5\linewidth]{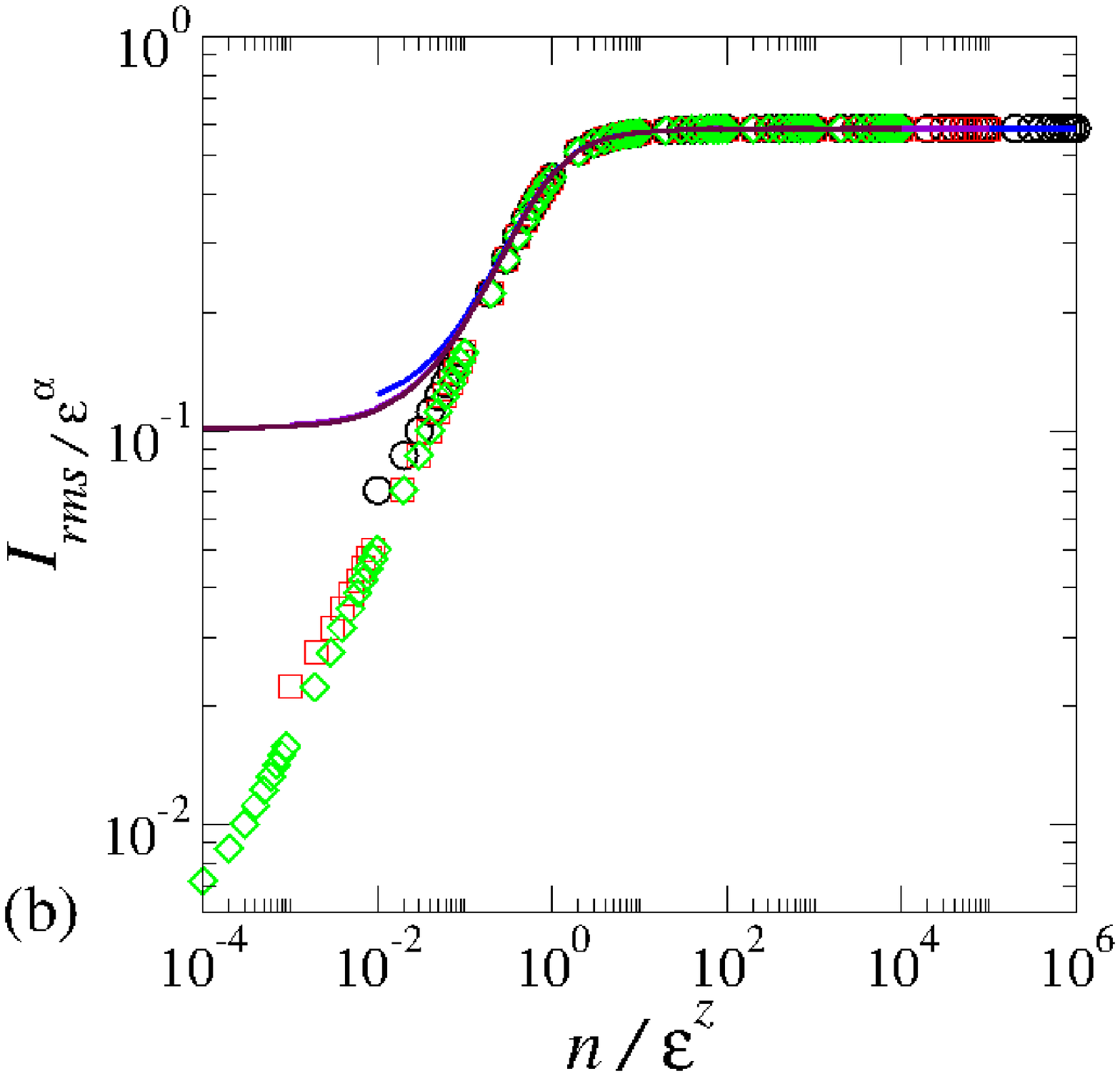}}
\caption{{(a) Plot of $I_{rms}~vs.~n$ for different control parameters 
as well as different values of $I_0$, as labeled in the figure. Symbols 
are used for $I_0\cong 0$ while continuous line represent 
$0<I_0<{{I_{fisc}}\over{\sqrt{3}}}$. (b) Overlap of the curves shown in 
(a) onto an universal plot after the scaling transformations.}}
\label{Fig3}
\end{figure}

Let us now discuss the case (ii) with 
$0<I_0<{{I_{fisc}}\over{\sqrt{3}}}$. Figure \ref{Fig3}(a)
shows a plot of $I_{rms}~vs.~n$ for different values of $\epsilon$ as 
well as different initial conditions, as labeled in the figure. The 
plateaus for short $n$ are evident, confirming the additional scaling. 
The initial conditions generating the plateaus were chosen as 
$I_0^{\prime}=I_0\left({{\epsilon^{\prime}}\over{\epsilon}}\right)^{{1}
\over{1+\gamma}}$. Figure \ref{Fig3}(b) shows the overlap of the curves 
depicted in (a) onto an universal plot, where curves with 
$0<I_0<{{I_{fisc}}\over{\sqrt{3}}}$ overlap between them for short $n$, 
then they join the regime of growth and finally bend, all together, 
towards a regime of saturation for large enough $n$.

Keeping now the approximation with $I_0\ne 0$ from equation 
(\ref{coseno}), substituting it into equation (\ref{irms}), 
the first crossover is given when
\begin{equation}
{{1}\over{3}}-{{4}\over{\pi^2}}\left(1-{{1}\over{2}}
\left({{\pi I_0}\over{I_{fisc}}}\right)^2\right)\left(1-{{\pi^2Dn_x^{
\prime
}}\over{2I^2_{fisc}}}\right)=1.
\end{equation}
Doing the proper algebra and considering only the leading term for the 
Taylor expansion in $I_0/I_{fisc}$ we end up with
\begin{equation}
n_x^{\prime}\propto {{\pi^2}\over{3}}{{I_0^2}\over{\epsilon^2}},
\end{equation}
as obtained in Ref. \cite{p8} for $\gamma=1$.

\begin{figure}[t]
%\vspace*{-0.8cm}
\centerline{\includegraphics[width=1.0\linewidth]{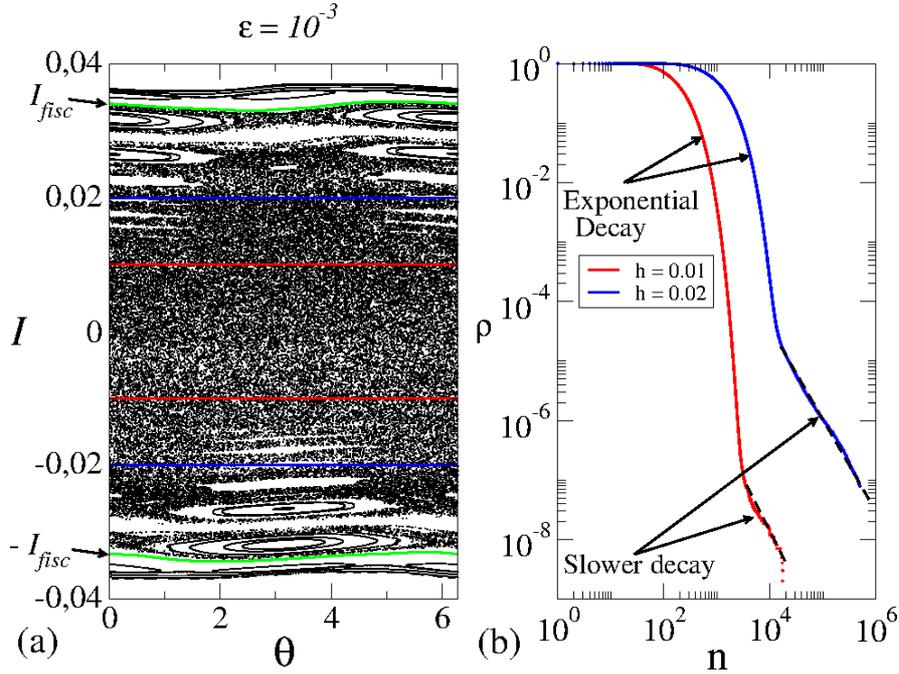}}
\caption{{(a) Plot of the phase space for mapping (\ref{eq4}) 
considering $\gamma=1$ and $\epsilon=10^{-3}$; (b) the behavior of the 
survival probability for orbits in the chaotic sea of (a) surviving the 
regions $h\in[-0.01,0.01]$ and $h\in[-0.02,0.02]$.}}
\label{Fig4}
\end{figure}

Let us now comment on the possible influence of the stickiness in the 
diffusion along the chaotic sea. As discussed in Ref. \cite{p12} and 
also in references therein, the survival probability corresponds to the 
probability a particle survive along the chaotic dynamics inside a 
given domain without escaping such region. The survival probability, 
is obtained from the integration of the escape frequency histogram 
written as
\begin{equation}
\rho(I_n)={{1}\over{N}}\sum_{i=1}^NN_{rec}(n),
\end{equation}
where $N$ denotes the number of different initial conditions, $n$ is 
the iteration number and $I$ is the action. We consider in our 
simulations to obtain $\rho$ an ensemble size of $N=10^9$ different 
initial conditions with $I_0=10^{-3}\epsilon$ and $\theta_0$ 
uniformly distributed in the interval $[0,2\pi]$ and a limiting number 
of iteration of $n=5\times 10^5$. We show the results for 
$\epsilon=10^{-3}$ although similar results would be obtained for other 
values of $\epsilon$. We define a given position along the action axis 
inside of the chaotic sea identified as $-I_{fisc}<h<I_{fisc}$. Starting 
the dynamics the orbit can evolve in the chaotic sea. If the particle
reaches $h$ we consider it escaped such region. We determine then the 
number of iterations spent until that time and started a different 
initial condition. This procedure is repeated until the ensemble of 
$10^9$ different phases is completely exhausted. If the phase space 
shows only chaos the survival probability is described by an exponential 
decay \cite{p12} while the existence of islands lead to local trapping, 
changing the exponential decay to a slower decay that might be a 
stretched exponential or a power law. Figure \ref{Fig4}(a) shows
a plot of the phase space for the parameter $\epsilon=10^{-3}$ and two 
regions delimited by the values $h=0.01$ (inside the red lines) and 
$h=0.02$ (inside the blue lines). The two curves shown in Fig. 
\ref{Fig4}(b) correspond to the survival probability obtained for 
chaotic orbits started in the chaotic sea with $I_0=10^{-3}\epsilon$ 
and different initial phases. We see they decay to start with at an 
exponential shape and eventually a small fraction of them gets stick 
near periodic islands for a while. The slower than exponential decay 
is a confirmation of the local and temporally confinement. The red 
curve decaying first corresponds to the region of $-0.01<h<0.01$. The 
periodic regions, not visible at the scale of the figure, influence only 
a portion of about $100$ particles from an ensemble of $10^9$, 
therefore statistically unidentifiable. The blue curve corresponds to a 
region of $-0.02<h<0.02$ and decays latter that the red curve. The 
stickiness affects about $50,000$ particles from an ensemble of $10^9$, 
therefore also very difficult to be detected in the simulations. Even 
though the stickiness is present at the domain investigated, its 
influence does not change the main results obtained.

\section{Discussions and conclusions}
\label{sec5}

We have investigated some scaling properties for a family of 
two-dimensional, nonlinear and area preserving maps whose angle diverges 
in the limit of vanishingly action. The parameter $\epsilon$ controls 
the nonlinearity of the problem and also a transition from integrable to 
non-integrable. The relevant scalings from the first momenta of the 
variable action is obtained numerically as well as by solution of the 
diffusion equation. For short $n$ and starting with low initial action, 
the behavior of $I_{rms}$ grows as $n$ with a power of $\beta=1/2$. 
Such exponent emerged naturally from the analytical solution. The 
regime of saturation scales with $\epsilon^{\alpha}$. Our finding for 
the regime of $\lim_{n\rightarrow\infty}$ yielded $I_{sat}\propto 
\epsilon^{{1}\over{(1+\gamma)}}$ in total agreement with the results 
already known in the literature \cite{p7,p8,p9,p11}. Our results 
obtained from the solution of the diffusion equation for the regime 
of initial action $0<I_0<{{I_{fisc}}\over{\sqrt{3}}}$ produces the 
additional plateau with the same relevant scaling as observed in Refs. 
\cite{p8,p11}. The additional scaling is related to the symmetry of the 
probability distribution function \cite{p9} and the break of symmetry 
produces the new crossover time.

The formalism presented in this paper has shown to be very robust and 
reliable to investigate chaotic transport. Future applications could 
involve study of diffusion of energy in time-dependent billiards, the 
scalings observed in the transition from limited to unlimited energy 
in a domain ranging from non-conservative to conservative system and 
many other applications. A problem that is still open and deserves much 
investigation relates to the dynamical trapping also called as 
stickiness for regions of the phase space near the invariant spanning 
curves. When a particle passes enough close such regions, they may stay 
trapped for a while - eventually long - until escape and run its own 
dynamics. During the trapping, the normal diffusion is mostly replaced 
by anomalous diffusion. This transition is not clear yet and deserves 
more investigation.

\begin{acknowledgements}
EDL acknowledges support from CNPq (303707/2015-1), FAPESP 
(2012/23688-5), (2017/14414-2) and FUNDUNESP. CMK thanks to CAPES for 
support.
\end{acknowledgements}

% BibTeX users please use one of
%\bibliographystyle{spbasic}      % basic style, author-year citations
%\bibliographystyle{spmpsci}      % mathematics and physical sciences
%\bibliographystyle{spphys}       % APS-like style for physics
%\bibliography{}   % name your BibTeX data base

% Non-BibTeX users please use

\end{document}